\documentclass[conference]{IEEEtran}

\usepackage[latin1]{inputenc}
\usepackage[T1]{fontenc}
\usepackage{fancyvrb}
\usepackage{times}
\usepackage{amsmath,amssymb}
\usepackage{euler}
\usepackage{url}
\usepackage{graphicx}
\usepackage{framed}
\usepackage[usenames,dvipsnames]{xcolor}
\usepackage{listings}

\newcommand{\secref}[1]{$\S$~\ref{#1}}

\begin{document}

\title{Eliminating Network Protocol Vulnerabilities Through Abstraction and
Systems Language Design}

\author{\IEEEauthorblockN{C. Jasson Casey\IEEEauthorrefmark{1},
Andrew Sutton\IEEEauthorrefmark{2},
Gabriel Dos Reis\IEEEauthorrefmark{2},
Alex Sprintson\IEEEauthorrefmark{1}}
\IEEEauthorblockA{\IEEEauthorrefmark{1}Department of Electrical and Computer Engineering,
Texas A\&M University}
\IEEEauthorblockA{\IEEEauthorrefmark{2}Department of Computer Science,
Texas A\&M University}}

\maketitle

\begin{abstract}
  Incorrect implementations of network protocol message specifications affect the
stability, security, and cost of network system development. Most implementation
defects fall into one of three categories of well defined message constraints.
However, the general process of constructing network protocol stacks and systems
does not capture these categorical constraints. We introduce a systems 
programming language with new abstractions that capture these constraints.
Safe and efficient implementations of standard message handling operations are 
synthesized by our compiler, and whole-program analysis is used to ensure 
constraints are never violated. We present language examples using the OpenFlow
protocol.

\end{abstract}

\section{Introduction}
\label{intro}

The message handling layer of any network protocol is notoriously difficult to
implement correctly. Common errors include: accepting or allowing the creation
of malformed messages, using incorrect byte ordering or byte alignment, using 
undefined values, etc. (\secref{vuln}). These defects lead to problems in 
stability, security, performance, and cost for network systems. 
Table~\ref{table:msg_tbl}, the result of a survey of the US-CERT Vulnerability 
Database~\cite{us_cert}, demonstrates that sophisticated organizations implementing 
mature protocols commit these errors. The persistent introduction of these 
defects is not the sign of an engineering problem, but a failure to use the 
correct levels of abstraction when working with network protocols.

\let\thefootnote\relax\footnotetext{This material is based upon work partially
supported by by the AFOSR under contract  \mbox{No. FA9550-13-1-0008}, and by
the NSF under grants  CCF-1150055 and ASI-1148461.}

\let\thefootnote\relax\footnotetext{978-1-4799-1270-4/13/\$31.00 \copyright2013 IEEE}


Message handling has been the focus of several research efforts.
When the wire-format of the message is not important, serialization solutions 
can be used~\cite{asn, proto_buf}. However, with network protocols, because of 
interoperability requirements, adherence to the specific wire-format is necessary.
As a result, a 
series of Domain Specific Languages (DSLs) that allow programmer control over 
the wire-format have been designed. These approaches synthesize data structures 
to hold messages and the typical operations necessary to manipulate them using 
correct by construction techniques~\cite{packet_types, datascript, binpac}. 
Language researchers have improved upon these DSLs with rich type systems that 
can prove certain safety properties, and address some of the problems mentioned 
previously~\cite{pads_orig, next_700_ddl}. Other work developed static analysis
techniques, that require no domain knowledge, to survey existing code bases and 
find occurrences of some of the previously mentioned defects~\cite{deputy}.

Systematically eliminating the categories of message related defects requires 
rich type systems and whole program analysis, which is not supported by existing
declarative DSLs. Invariants and semantic information produced by the DSL is not
incorporated or used in program analysis by the target language. Furthermore, 
finding all occurrences of message related defects require some level of domain
knowledge during static analysis. This must either be built in to the language 
or programmer specified in a way that resembles existing network protocol 
specifications. Analysis by formal methods should be a by-product of compiling 
the network program, and not require any specialized knowledge by the 
programmer. Our work-in-progress develops a systems programming language to 
address these issues. This allows for full program analysis, providing stronger 
safety guarantees and offering domain specific optimization that is exceedingly 
difficult to accomplish by hand or impossible with a DSL.

\begin{table}[b]
\vspace{-12pt}
\caption{Message Related Vulnerabilities}
\begin{tabular}{|c|c|c|c|c|c|}
  \hline
  Proto. & Age & Bug Date & Vendor & Error & CERT \# \\ \hline
  \hline
  802.11i & 2004 & 2012 & Broadcom & semantic & 160027 \\ \hline
  OSPFv2 & 1998 & 2012 & Quagga & struct & 551715 \\ \hline
  NTPD & 1985 & 2009 & GNU & struct & 853097 \\ \hline
  ICMP & 1981 & 2007 & Cisco & both & 341288 \\ \hline
  VTP & 1996 & 2006 & Cisco & semantic & 821420 \\ \hline
  Bootp & 1985 & 2006 & Apple & struct & 776628 \\ \hline
\end{tabular} \\
\label{table:msg_tbl}
\end{table}

In this paper we clearly identify categories of message related vulnerabilities
by their structural and semantic constraints. We show that these
categories are responsible for known vulnerabilities, and using our tools we 
show that even some live Internet traffic violates these constraints. We then 
introduce a systems programming language that allows programmers to capture 
network protocol message structure and constraints. The constraints allow the 
compiler to reason over entire programs, identifying and eliminating the 
categories of vulnerabilities mentioned before. Additionally, the choice of a 
systems programming language allows for efficient code generation. 
Throughout the paper we use OpenFlow~\cite{ofp} as our reference protocol. 
Our contributions are:
\vspace{-2pt}
\begin{itemize}
  \item the identificaiton of three categories of network protocol 
        vulnerabilities common in network programs and in network traffic 
        (in \secref{vuln}),
  \item the development of abstractions that prevent the construction of
        messages that lead to these vulnerabilities, and unsafe access of 
        conditional fields (in \secref{vuln} and \secref{lang}),
  \item the design of a systems programming language that eliminates
        unsafe protocol implementations through type checking (in \secref{lang}, 
        \secref{compiler}, and \secref{safety_opt}), and
  \item the implementation of a compiler and library supporting the language
        (in \secref{impl_eval}).
\end{itemize}

\section{Message Vulnerabilities}
\label{vuln}

There are three categories of message vulnerabilities that we address: 
structural constraint violation, semantic constraint violation, and unsafe
access of conditional fields. In this section we describe these
categories in detail with examples using the OpenFlow v1.0 protocol~\cite{ofp}.
We first briefly describe the protocol and then show vulnerability examples.

\begin{figure}[h]
\vspace{-12pt}
\begin{center}
  \includegraphics[width=.4\textwidth]{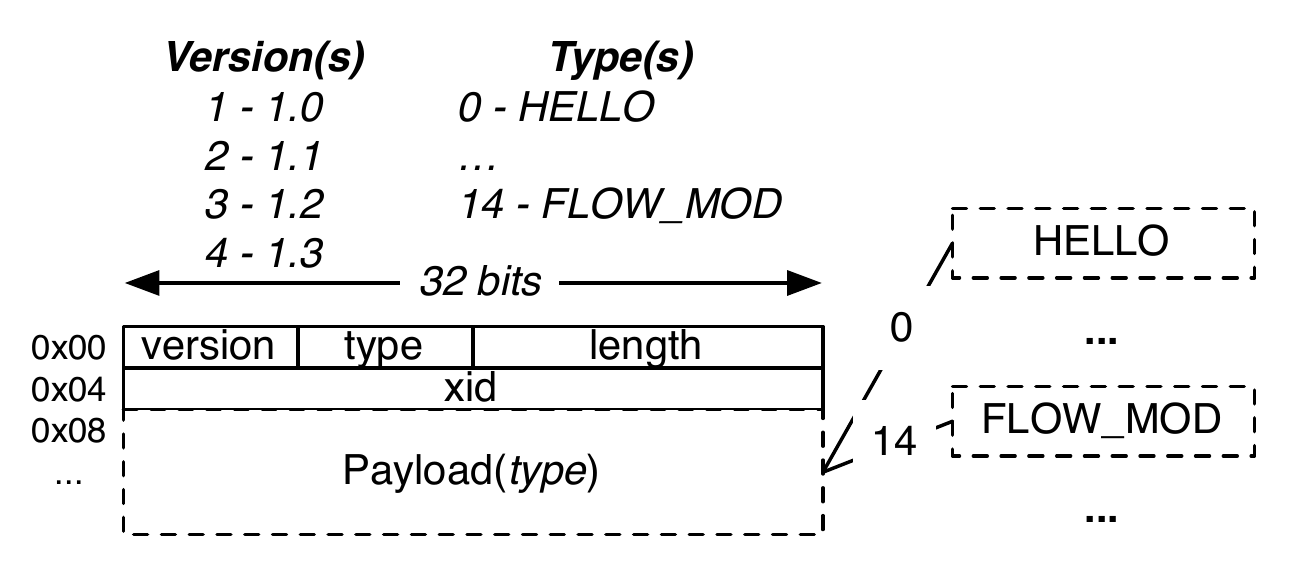}
  \vspace{-6pt}
  \caption{OpenFlow Message Format}
  \label{figure:ofp_msg_rep}
\end{center}
\vspace{-12pt}
\end{figure}

Figure~\ref{figure:ofp_msg_rep} summarizes the message format, or
representation, of an OpenFlow message. A message consists of a fixed 8 byte
header followed by a variable sized payload. The header indicates the version of
the OpenFlow protocol, the type of the payload, the length of the entire
message, and a transaction identifier used to match response messages to their
requests. The payload can be one of 22 types in version 1.0. This protocol was
designed to operate over stream-oriented transports, which are more difficult to
handle than datagram or message oriented transports. Streams have no concept of
message boundaries and can give the program anything from a single byte to
several messages in a single read. It is the programmer's responsibility, not
the transport's, to determine where the payload ends and the next message
begins.

\textbf{Semantic constraints} ensure that a message field's value has a defined 
meaning in the protocol. For instance, in OpenFlow 1.0 the domain of the version field is
$[1,1]$, the domain of the type field is $[0,21]$, the domain of the length 
field is $[8,2^{16}-1]$, and the domain of the xid field is $[0,2^{32}-1]$. Any 
value that is not in the domain of its field is semantically invalid; it has no
meaning in the protocol definition. Violating a semantic constraint is similar
to using undefined behavior in a programming language. As 
Table~\ref{table:msg_tbl} shows, constraint violations lead to vulnerabilities.

\textbf{Structural constraints} address how messages are constructed and used 
by a program. Messages are constructed in two ways: either by a program to send 
over a network, or from the network to send to the program. In both cases it is 
important to construct only structurally valid messages. 

These constraints deal with the number of bytes a message occupies in
buffers used to communicate with the network. 
Structural constraint testing is the process of ensuring there are enough bytes 
to complete the operation of constructing a message from a buffer, or filling a
buffer with a well-formed message.
For example, any buffer containing less than 8 bytes cannot possibly represent 
a valid OpenFlow message and any attempt to interpret that buffer as a
message, would be an error. 

Semantic constraint violations can lead to structural constraint violations. 
Reading from a stream can produce a buffer containing several 
messages. The header of the first message must be contained in the first 8 
bytes, and the payload must end at the position where the header's length field 
indicates. In the header, the length field can be semantically invalid, but
because it is used to constrain the payload's size, it becomes a structural 
constraint violation as well. A similar problem arises with a semantically 
invalid type field that is used to choose the payload.

\textbf{Safe access} ensures that fields with run-time dependency are always
validated before use. Many fields in OpenFlow are dependent; their meaning
is determined by the values of previously encountered fields. For 
example, there is a dependency between the payload of a message and the header's
type field. A structurally and semantically valid Hello message cannot have its
payload treated as the FlowMod type without invoking undesired behavior.

\section{Language}
\label{lang}

This work builds on fundamental notions from system
programming~\cite{iso03,eop}, from structured generic programming, and
mathematical programming languages such as AXIOM~\cite{jenks92}, and
Liz~\cite{liz}. This work, heavily  inspired by the Liz language, aims to
support simple, safe, and efficient  handling of network protocol messages. The
core of our language supports values, references, constants, functions, records,
a minimal set of expressions, and  follows Call-By-Value semantics. We do not
expose pointers to the users of the  language and drastically restrict heap
allocation to certain language built-in types. 
Figure~\ref{figure:syntax} shows the abstract syntax of the language. The 
dependent types $\omega$, are a primary contribution of this paper.

\begin{figure}
\begin{center}
  \begin{framed}
    \input{syntax}
  \end{framed}
  \vspace{-6pt}
  \caption{Core Language Syntax}
  \label{figure:syntax}
\end{center}
\vspace{-18pt}
\end{figure}

Our language captures the structural and semantic constraints of a message
with user-defined type and variable declarations. We then enforce these 
constraints through the process of object and symbolic construction. If object
construction completes successfully, then structural constraints are upheld, if
symbolic construction completes successfully, then semantic constraints hold.
Using construction to establish invariants is a common way to reason about
program behavior. 

An object, or instance of a type, must be constructed before use. The process of
object construction involves allocating space where the object will live, and
initializing its values to establish its invariant. Symbolic construction 
extends object construction to include ensuring the value of the object is 
consistent with its symbolic constructor. Upon completion of construction, an 
object is well-formed and its invariant has been established.

\textbf{The $\omega$ types}, see Figure~\ref{figure:syntax}, allow user 
definition of precise structural and semantic constraints. Structural 
constraints can be explicitly stated by the specifiers: $bits(e)$, and 
$constraint(e)$. Structural constraints are otherwise implicit in the type, which will be explained
below. Semantic constraints are introduced through the declaration syntax. Any 
declaration can impose a semantic constraint through the use of the bar, $|$, 
followed by a guard expression. These declarations provide constraint 
information to the compiler that is used in full program analysis. This allows 
the compiler to reason about both constraint satisfaction and safe usage. The
following is a short summary of the $\omega$ types:
\begin{itemize}
  \item $uint(spec[,xform])$ defines an unsigned integer that has a precise bit
  width specified by its structural constraint ($spec$). Optionally, the
  type also takes a transform parameter ($xform$) that allows for non-native
  representation of the data. For instance, a protocol may specify that a value's
  representation is Most Significant Byte First (MSBF) and 1's complement. 
  \item $array(\tau,e)$ and $vector(\tau)$ types allow for sequences of objects
  of type $\tau$.
  The array is statically sized to contain $e$ elements, while the vector has a 
  dynamic size.
  \item $record\{\overline{decl}\}$ is a sequence of declarations, whose objects
  can be accessed by field name. There is no padding, alignment, or
  meta-data applied to the object by the compiler. If padding,
  alignment, or meta-data is desired, the programmer can explicitly declare them
  as fields. Additionally, the order of fields is preserved.
  \item $variant\{\overline{\tau \mbox{ if } e}\}$ is a union of types where 
  each type is guarded by a predicate. A variant is constructed by evaluating
  the predicate set and invoking the constructor of the corresponding true
  predicate. A variant is uninitialized until evaluation of the predicate set.
\end{itemize}

\begin{figure}[h]
\begin{center}
\begin{lstlisting}
define Hdr:type = record {
  vrsn: uint(bits(8)); 
  type: uint(bits(8));
  len:   uint(bits(16), msbf);
  xid:   uint(bits(32), msbf);
}
define Pld(x:uint):type = variant {
  Hello    if x == 0;
  ...
  FlowMod if x == 14;
  ...
}
define valid_hdr(h:cref(Hdr)):bool = {
  return h.vrsn == 1
    and h.len >= bytes(h);
}
define Msg:type = record {
  hdr: Hdr | valid_hdr;
  pld: Pld(hdr.type, 
       constraint(hdr.len - bytes(hdr)));
}
\end{lstlisting}
  \vspace{-6pt}
  \caption{OpenFlow v1.0 Message Declaration}
  \label{figure:msg}
\end{center}
\vspace{-18pt}
\end{figure}

Figure~\ref{figure:msg} demonstrates how to declare types corresponding to the
OpenFlow header, payload, and message. The header is a simple record of four 
fields, all of which have constant specifiers, and follow MSBF ordering. The
payload type is a unique choice of types based on a type parameter. For
the values in the header to have semantic meaning in version 1.0 of the OpenFlow
protocol we have to constrain their values. This is achieved with the semantic 
constraint $valid\_hdr$, which must first be defined as a function that takes
a constant reference to a header and returns bool. Message is defined as a 
record including a header with version 1.0 semantic constraints, and a payload 
that is parameterized over the header's type field and constraint. All type
constructors, with the exception of $uint$ and $array$ allow for an optional
specifier constraint. In this particular case, construction of Pld is not to 
exceed the result of the constraint $hdr.len - bytes(hdr)$.

\textbf{Buffer \& View} are abstractions over the underlying machine architecture
that help the compiler ensure structural constraints are never violated. Reading
from a file or socket results in a buffer with begin and end boundaries 
surrounding the bytes received. $View$ is a mechanism that restricts visibility
into a $Buffer$. The set of operations defined for $Buffer$ and $View$ are:
\begin{itemize}
  \item $view$: returns a view of an entire buffer
  \item $available$: returns the byte size of a view
  \item $advance$: returns a view with an advanced head
  \item $constrain$: returns a view with a constrained tail
  \item $put$: writes a value to a view
  \item $get$: reads a value from a view
\end{itemize}

\begin{figure}[h]
\vspace{-12pt}
\begin{center}
  \includegraphics[width=.3\textwidth]{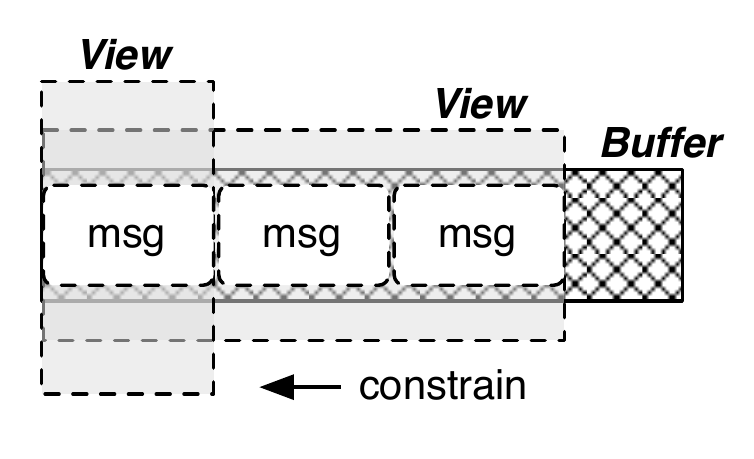}
  \vspace{-12pt}
  \caption{Constraining a View limits the number of bytes available for access}
  \label{figure:view_ops}
\end{center}
\vspace{-6pt}
\end{figure}

Figure~\ref{figure:view_ops} illustrates a buffer returned from a read
system call against a TCP socket. A single read has resulted in more than one 
protocol message. The initial view wraps all of the data; however, the first
few bytes of the view contain a protocol header, which provides the length of
the first message. This length is then used to constrain the visibility to
precisely one message. The constrain operation supports use of datagram and
stream oriented transports, while also providing safety boundaries for object
construction. 

\section{Compiler Synthesis}
\label{compiler}

Programmers continue to make mistakes implementing message operations that are 
necessary in all protocols such as those mentioned in \secref{intro} and 
\secref{impl_eval}. 
Our strategy is to 
eliminate the need to write these common operations by having the compiler 
synthesize safe and efficient versions. All user-defined type definitions
contain structural and semantic constraints. This information is sufficient to
synthesize the following operations:
\begin{itemize}
  \item $construction$: constructs an object from expressions
  \item $copy\_construction$: constructs an object from another
  \item $assignment$: copies an object's state into another
  \item $bytes$: returns the number of bytes of an object
  \item $to\_view$: writes an object to a view
  \item $from\_view$: constructs an object from a view
  \item $equal$, $not\_equal$: compare objects for equivalence
  \item $to\_string$: returns a string representation of the object
\end{itemize}

The remainder of this section will describe the synthesis process for a small 
subset of the above operations focusing on synthesis and constraint validation
for $bytes$ and $from\_view$.

\textbf{Bytes} is the name of the operation for determining the byte size of any
object. The operation $bytes$ is synthesized for each declared $\omega$ type in 
a program. For $uint$ it returns the number of bytes indicated by the specifier.
For $array$ it returns the result of $sizeof(T) * elements$, where $T$ is the
type contained by the $array$. For objects of type $Vector$, bytes is not a 
constant expression. It has run-time dependencies and returns the accumulation 
of calling $bytes$ over its elements. $Record$ returns the sum of calling 
$bytes$ over its constituent fields, and is only a constant if all of its fields
are also constant. Calling $bytes$ over a $variant$ returns 0 if the variant is
uninitialized, or it proxies the call to the contained object. 
Figure~\ref{figure:synth_bytes} illustrates the process just described in pseudo
code. Several other synthesized operations depend on $bytes$.

\textbf{Structural constraint violations} must be prevented during object
construction. There are only three ways to violate a structural 
constraint: overflowing a view, underflowing a view, or constructing an invalid 
variant. Overflowing a view involves advancing the view beyond the number of 
bytes contained. Underflowing a view involves constraining a view by more bytes 
than contained. Constructing an invalid variant is caused when none of the 
variant's contained type predicates evaluate to true.

In order to construct the OpenFlow header the size of the view must be at least
as large as the number of bytes of the header, or 8 bytes. During object 
construction the view is always advanced by the size in $bytes$ of the object.
Constructing the message object from a view of 7 or less bytes would result in a
view overflow.

The OpenFlow protocol indicates the length of a message with the header length
field. This value is used to constrain the view for payload construction. It is
possible, through accident or malicious intent, for the length field to be
inconsistent with the amount of data actually sent. If the field indicated less
than 8 bytes of payload it could be possible to underflow the view.

The value used to construct the variant payload is in the header. Again, either
through accident or malicious intent, it is possible for the header to indicate
a type which will not result in valid variant construction. Using a variant in
an invalid way will result in undefined behavior.

\begin{figure}[t]
\begin{center}
\begin{lstlisting}
define bytes(u : uint(bits(v), f)) : uint = {
  return v/8;
}
define bytes(a : cref array(T,x)) : uint = {
  return sizeof(T) * x;
}
define bytes(v : cref vector(T) : uint = {
  accum : uint = 0;
  foreach(item in v) {
    accum += bytes(v);
  }
  return accum;
}
define bytes(r : cref record) : uint = {
  return bytes(r.x1)+bytes(r.x2)+...+bytes(r.xN);
}
define bytes(v : cref variant) : uint = {
  if(not init(v))
    return 0;
  switch(v.kind) {
    case variant::K1 => return bytes(T1(v.value));
    case variant::K2 => return bytes(T2(v.value));
    ...
    case variant::KN => return bytes(TN(v.value));
  }
}
\end{lstlisting}
  \vspace{-12pt}
  \caption{Synthesis rules for the $bytes$ operation}
  \label{figure:synth_bytes}
\end{center}
\vspace{-24pt}
\end{figure}

\textbf{Object construction} is possible with either a constructor that operates
over expressions or using the $from\_view$ operation. 
Figure~\ref{figure:synth_from_view} illustrates pseudo code for synthesizing 
$from\_view$. The operation returns false when a structural constraint has been 
violated. Failure indicates a partially constructed object. For simple 
types, such as $uint$ and $array$, the structural constraints are 
always checked. If there are not enough bytes in the view to complete the 
operation, the operation fails. Otherwise, the object's value is constructed by
reading from the view. If a $xform$ is present the object's value is updated
using the specified transform. Finally, the view is advanced by the size of the
object just constructed.

\begin{figure}[h]
\begin{center}
\begin{lstlisting}
define from_view(v:ref(view), u:uint(bits(v),f)):bool={
  if (available(v) < bytes(u))
    return false;
  get(v, u);
  u = f(u);
  advance(v, bytes(u));
  return true;
}
define from_view(v:ref(view), a:ref(array(T,e))):bool={
  if (available(v) < bytes(a))
    return false;
  foreach(x in a) {
    from_view(v, x); 
  }
  return true;
}
define from_view(v:ref(view), vc:ref(vector(T))):bool={
  while(available(v) > 0) {
    t:T;
    if (not from_view(v, t))
      return false;
    push(vc, t);
  }
  return true;
}
define from_view(v:ref(view), r:ref(record)):bool={
  if (not from_view(v, r.x1))
    return false;
  if (not from_view(v, r.x2))
    return false;
  ...
  return from_view(v, r.xN);
}
define from_view(v:ref(view), vr:ref(variant)):bool={
  if(not init(vr))
    return false;
  switch(vr.kind) {
    case vr::K1 => return from_view(v, T1(vr.value));
    case vr::K2 => return from_view(v, T2(vr.value));
    ...
    case vr::KN => return from_view(v, TN(vr.value));
  }
}

\end{lstlisting}
  \vspace{-6pt}
  \caption{Synthesis rules for the $from\_view$ operation}
  \label{figure:synth_from_view}
\end{center}
\end{figure}

The vector version of $from\_view$ operates in a greedy fashion, it will consume
the entire view. If this behavior is not desired the view must be constrained 
before construction. As long as there are bytes in the view the vector
will attempt to construct an object. Upon success, the object is inserted into 
the vector and the process repeats. The record version will attempt to 
construct its constituent fields and either return at the first failure 
or succeed. The variant must guard against the third type of structural 
constraint; it must be initialized to a valid type. If this is not true then 
$from\_view$ will fail, otherwise $from\_view$ is called over the appropriate
type.

\textbf{Symbolic construction} ensures that all run-time type dependencies must 
be propagated and semantic constraints are inserted into synthesized code. The
OpenFlow message from Figure~\ref{figure:msg} has semantic constraints, a 
run-time type dependency, and a constrained view. The semantic constraint turns 
into a predicate check immediately after the call to $from\_view$ of the header.
If the check fails, the operation immediately returns with failure. Next, the 
type parameter must be checked and initialized before the call to $from\_view$
over the payload. The check ensures that failure happens if the value is 
undefined, and if upon success initializes the payload's kind. Finally, the 
constrained view operation is propagated. The result of this final step is shown
in Figure~\ref{figure:from_view_msg}.

\begin{figure}[t]
\vspace{-12pt}
\hrule
\begin{center}
\begin{lstlisting}
define from_view(v:ref(view), m:Msg):bool = {
  if(not from_view(v,m.hdr))           return false;
  if(not valid_hdr(m.hdr))             return false;
  if(not construct(m.pld, m.hdr)) return false;
  return from_view(constrain(v,m.hdr.len-bytes(m.hdr)), 
                           m.pld);
}
\end{lstlisting}
\vspace{-6pt}
\caption{Synthesis of $from\_view$ for Msg}
\label{figure:from_view_msg}
\end{center}
\vspace{-18pt}
\end{figure}

\section{Safety And Optimization}
\label{safety_opt}

There are only three ways to violate structural constraints: view underflow,
view overflow, and reading or writing from an uninitialized variant. These three
categories of mistakes can be identified with two simple invariants. If the
bytes of a view are always non-negative then it is impossible to underflow or
overflow the view. If a read or write of a variant is always preceded by a valid
initialization then the third category is also impossible.
The compiler uses a dataflow analysis framework to prove that these invariants
always hold or fail the compilation of the program with useful error messages.

The goal of the compiler is to synthesize safe and efficient code. The 
synthesis algorithm described previously will produce safe but inefficient 
code. The operations $to\_view$ and $from\_view$ contain fine-grained guards 
that protect against view underflow and overflow. These fine-grained guards are
the source of inefficiency. For example, it is better to have a single guard
over a sequence of objects of constant size, then to have a guard over each 
object. We call this guard fusing, it is analogous to fusing basic blocks to
form larger basic blocks. It is also useful to lift guards from inside the
called function to the call site. By lifting guards to the call site, potential
guard fusing optimization becomes possible. The optimization strategy that we
follow is a set of fusing and lifting rounds that reduce the number of guards
and form the largest possible object construction basic blocks for $to\_view$
and $from\_view$.

\begin{figure}[h]
\vspace{-12pt}
\begin{center}
  \includegraphics[width=.4\textwidth]{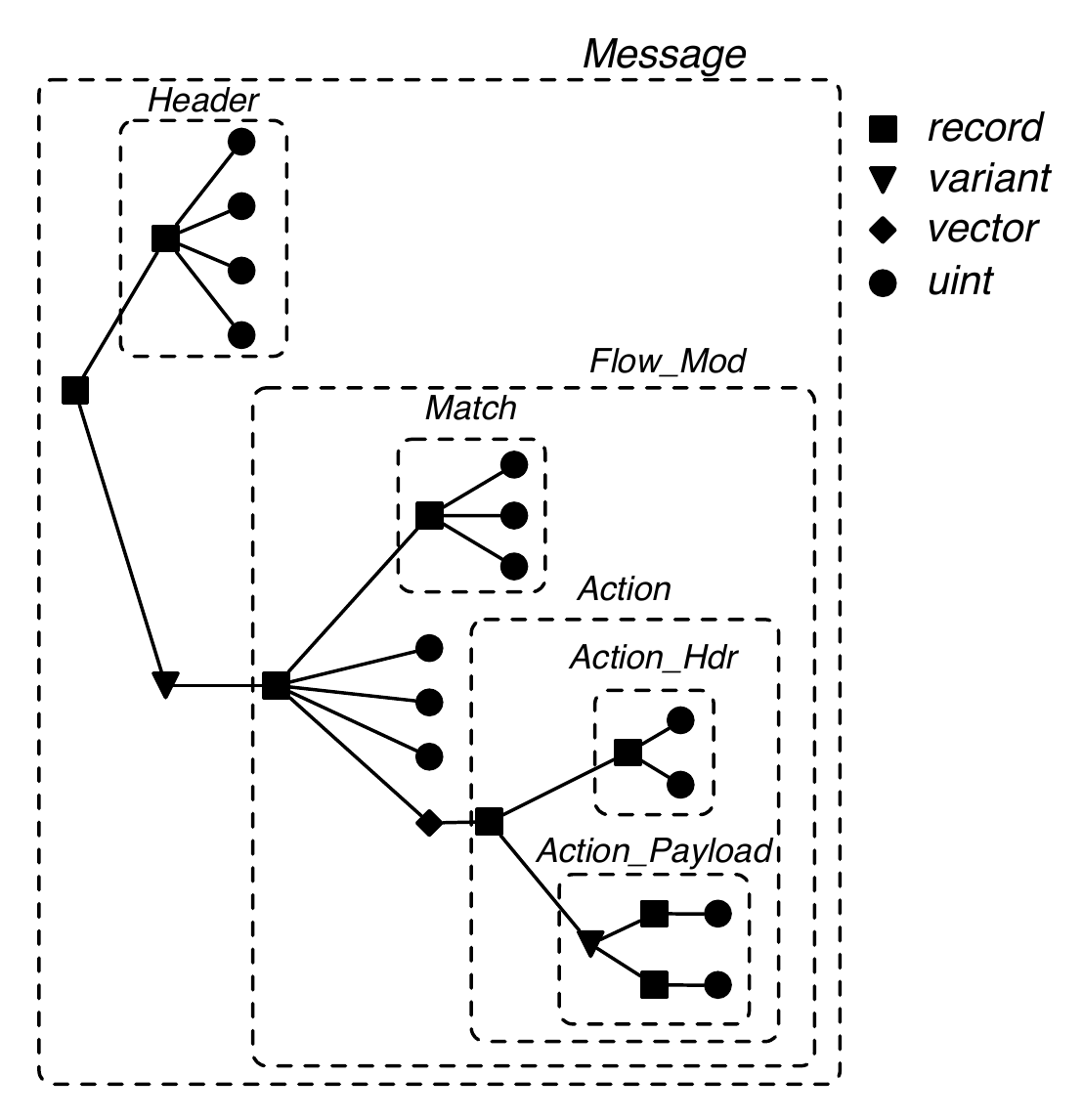}
  \vspace{-6pt}
  \caption{Flow Modification Control Flow Graph}
  \label{figure:flow_mod_deps}
\end{center}
\vspace{-12pt}
\end{figure}

All synthesized operations have a similar Call Sequence Graph (CSG).
Figure~\ref{figure:flow_mod_deps} illustrates the generalized CSG for a
Flow Modification message. Each node represents a function in the call
sequence, the function types are indicated by the node shape in the figure. This
CSG is used to both synthesize operation definitions and analyze safe usage of
messages. Guards start at the leaves of the graph and are lifted to their parent
nodes. If all guards can be fused within an interior node, then the new guard is
lifted and the process is repeated. This process ensures guards covering the 
largest possible constant sized objects are performed, additionally this process
is unaware of
protocols and will optimize across layer boundaries. Sometimes a node can have
more than one parent node, where the parents have differing behaviors. In this 
case, we split the node into two versions where we lift the guard when it is
contained within a constant structure, or leave it in place otherwise.

\textbf{Code generation} takes place after the optimization phase. We currently
support C++11 as our target language. All message related type definitions and 
synthesized operations will be written to a single set of .cpp and .hpp files, 
the program itself is written to main.cpp. 

\section{Implementation and Evaluation}
\label{impl_eval}
Protocol implementations that send messages that are either structurally or
semantically invalid exist. Applying this work we were able to discover
structural constraint violations in packet traces from core Internet routers.
Furthermore, we were able to define three categories of message constraints, and
show violation of these constraints can lead to high profile vulnerabilities.
Network programs must always handle messages in a safe manner. To this aim, we 
developed a systems programming language and library for writing safe efficient
network programs.

\textbf{The language implementation} was originally developed as a C++11 
library. The library was used to test ideas and guide the language design.
However, in order to enforce safety guarantees with optimization a compiler was
necessary. We experimented with the language using two types of network 
programs: a protocol analyzer, and an OpenFlow stack. This set of applications
provided good coverage over the diversity of protocol formats and their
constraints.

\textbf{Core Internet traces} were obtained from Caida~\cite{caida} as
test data for packet analyzers written in our new language facilities. The
traffic was recorded from high speed interfaces, OC-192 (up to 10Gbps). Only
the Layer 3 and Layer 4 headers, timing, and summary information are present.
The layer 3 addresses were randomized and the layer 4 payloads were removed for
anonymization purposes. Each trace has between 500 MB and 1 GB of data and
was timestamped in minute intervals. We analyzed a 10 minute segment of traces
from October of 2012. 

We focused on looking for structural and value constraints violations 
within IPv4, IPv6, TCP, and UDP. Table~\ref{table:caida} shows that we found
structural constraint violations in all but one protocol; no semantic
constraint violations were found. IPv4 violated its structural constraint
with regards to IPv4 Options. The values in the Internet Header Length (IHL)
field indicated a number of Options that should be constructed; however, this
packet would overflow its view, the received block of data was too small. 
The TCP and UDP structural constraint violations were of the same nature; they 
violated the basic constraint of a minimum sized header.

\begin{table}[h]
\begin{center}
\caption{Caida Traces \label{table:caida}}
\begin{tabular}{|c|c|c|c|c|}
   \hline
   Desc. & IPv4 & IPv6 & TCP & UDP \# \\ \hline
   \hline
   Count & 247,849,217& 130,760 & 221,243,574 & 23,633,921 \\ \hline
   CDF & 99.95\% & 0.05\% & 89.22\% & 9.53\% \\ \hline
   Struct & 16 & 0 & 84,274 & 86,123 \\ \hline
\end{tabular} \\
\end{center}
\vspace{-6pt}
\end{table}

The source of these structural constraint violations is not currently known. It
could be evidence of unintentional errors in sending devices, it could be 
maliciously crafted packets, or it could be due to the collection process of
the trace data. However, regardless of the source, structural constraints have
been violated and these packets should not be admitted to safe network programs.

\vfill

\section{Conclusion And Future Work}

Incorrect implementations of protocol message specifications affect the
stability of network systems and potentially lead to vulnerabilities. In this
paper we identified three categories of constraints that can be used either to
test whether a message is well-formed or to generate safe code. We developed a
systems programming language that allowed user-defined types to capture these
constraints as well as a reasoning framework to ensure these constraints are
always upheld within the users program. We presented example type definitions 
and compiler synthesized code using the OpenFlow 1.0 protocol.

The next steps for this work fall into two categories: extending the $\omega$
types and formalizing the meta-system. Extending the type system will allow for
the support of more protocols. Vectors will be extended to support termination
predicates as a structural constraint parameter, this will allow for 
self-terminating sequences such as null-terminated character strings. 
Generalized enumerations will be added as an easier mechanism for restricting
values used in message construction. Finally, work on the meta-system is focused
on generating proof certificates that can be used for mechanical verification of
safety.

\bibliographystyle{plain}
\bibliography{references}

\begin{thebibliography}{10}

\bibitem{caida}
The {CAIDA Anonymized 2012 Internet Traces} - 20120119-125903, {kc claffy, Dan
  Andersen, Paul Hick}.

\bibitem{datascript}
G.~Back.
\newblock Datascript- a specification and scripting language for binary data,
  2002.

\bibitem{deputy}
J.~Condit, M.~Harren, Z.~Anderson, D.~Gay, and G.~Necula.
\newblock Dependent types for low-level programming.
\newblock {\em Programming Languages and Systems}, pages 520--535, 2007.

\bibitem{pads_orig}
K.~Fisher and R.~Gruber.
\newblock {PADS}: a domain-specific language for processing ad hoc data.
\newblock {\em SIGPLAN Not.}, 40(6):295--304, June 2005.

\bibitem{next_700_ddl}
K.~Fisher, Y.~Mandelbaum, and D.~Walker.
\newblock The next 700 data description languages.
\newblock {\em SIGPLAN Not.}, 41(1):2--15, January 2006.

\bibitem{proto_buf}
Google.
\newblock Protocol buffers, 2012.
\newblock \url{http://code.google.com/p/protobuf/}.

\bibitem{us_cert}
US~Government.
\newblock {United States Computer Emergency Readiness Team}, 2012.

\bibitem{iso03}
International Organization for Standards.
\newblock {\em {International Standard ISO/IEC 14882. Programming Languages ---
  C++}}, 2nd edition, 2003.

\bibitem{jenks92}
R.~D. Jenks and R.~S. Sutor.
\newblock {\em {AXIOM: The Scientific Computation System}}.
\newblock Springer-Verlag, 1992.

\bibitem{packet_types}
P.~J. McCann and S.~Chandra.
\newblock Packet types: abstract specification of network protocol messages.
\newblock In {\em Proceedings of the conference on Applications, Technologies,
  Architectures, and Protocols for Computer Communication}, SIGCOMM '00, pages
  321--333, New York, NY, USA, 2000. ACM.

\bibitem{ofp}
N.~McKeown, T.~Anderson, H.~Balakrishnan, G.~Parulkar, L.~Peterson, J.~Rexford,
  S.~Shenker, and J.~Turner.
\newblock Openflow: enabling innovation in campus networks.
\newblock {\em SIGCOMM Comput. Commun. Rev.}, 38(2):69--74, March 2008.

\bibitem{binpac}
R.~Pang, V.~Paxson, R.~Sommer, and L.~Peterson.
\newblock binpac: a yacc for writing application protocol parsers.
\newblock In {\em Proceedings of the 6th ACM SIGCOMM conference on Internet
  measurement}, IMC '06, pages 289--300, New York, NY, USA, 2006. ACM.

\bibitem{liz}
G.~Dos Reis.
\newblock {A System for Axiomatic Programming}.
\newblock In {\em AISC/MKM/Calculemus}, pages 295--309, 2012.

\bibitem{eop}
Alexander~A Stepanov and Paul McJones.
\newblock {\em Elements of programming}.
\newblock Addison-Wesley Professional, 2009.

\bibitem{asn}
International~Telecommunication Union.
\newblock Abstract syntax notation one ({ASN.1}).
\newblock Technical report, 2002.
\newblock Available from:
  http://www.itu.int/ITU-T/studygroups/com17/languages/X.680-0207.pdf.

\end{thebibliography}

\end{document}